\documentclass[journal]{IEEEtran}
\usepackage{amsmath,amsfonts}
\usepackage{algorithmic}
\usepackage{algorithm}
\usepackage{array}
\usepackage[caption=false,font=normalsize,labelfont=sf,textfont=sf]{subfig}
\usepackage{textcomp}
\usepackage{stfloats}
\usepackage{url}
\usepackage{verbatim}
\usepackage{graphicx}
\usepackage{braket}
\usepackage{cite}
\usepackage{color}
\usepackage{xcolor}
\usepackage{hyperref}
\usepackage{ragged2e}
\usepackage{bm}

\newcommand{\state}[1]{\ket{#1}}
\begin{document}

\title{Quantum Reinforcement Learning for 6G and Beyond Wireless Networks}

\author{Dinh-Hieu Tran, Thai Duong Nguyen, Thanh-Dao Nguyen, Ngoc-Tan Nguyen, Van Nhan Vo, Hung Tran, Mouhamad Chehaitly, Yan Kyaw Tun, Cedomir Stefanovic, Tu Ho Dac,  Eva  Lagunas, Symeon Chatzinotas, and Nguyen Van Huynh}

\maketitle

\begin{abstract}
While 5G is being deployed worldwide, 6G is receiving increasing attention from researchers to meet the growing demand for higher data rates, lower latency, higher density, and seamless communications worldwide. To meet the stringent requirements of 6G wireless communications networks, AI-integrated communications have become an indispensable part of supporting 6G systems with intelligence, automation, and big data training capabilities. However, traditional artificial intelligence (AI) systems are difficult to meet the stringent latency and high throughput requirements of 6G with limited resources. In this article, we summarize, analyze, discuss the potential, and benefits of Quantum Reinforcement Learning (QRL) in 6G. As an example, we show the superiority of QRL in dynamic spectrum access compared to the conventional Deep Reinforcement Learning (DRL) approach. In addition, we provide an overview of what DRL has accomplished in 6G and its challenges and limitations. From there, we introduce QRL and potential research directions that should continue to be of interest in 6G. To the best of our knowledge, this is the first review and vision article on QRL for 6G wireless communication networks.

\end{abstract}


\section{Introduction}

While 5G is being deployed and operated worldwide, new research is still being done towards 6G to meet the growing demand for higher data transmission, lower latency, and wider coverage worldwide. 6G is expected to be released in the early 2030s, and many challenges need to be solved before then. 6G promises to bring drastic changes compared to 5G, such as transmission speeds 50 to 100 times faster than 5G, using a wider frequency range including sub-THz bands (e.g., 90 GHz to 3 THz in 6G) or new mid-band spectrum from 7 to 15 GHz, thus providing much larger bandwidth than 5G. In addition, 6G also offers significantly lower latency than 5G, e.g., it expects to reduce the transmission delay to 0.1 ms compared to 1 ms in 5G. In addition, 6G will provide infrastructure for a much larger number of mobile users (MUs) than 5G, thus meeting the widespread connectivity needs of massive Internet of Things (IoT) devices. In addition, 6G promises to enable new applications such as holographic, extended reality, haptics, AI integration, advanced medical applications, and Metaverse communications.

Along with the potential promises discussed above, the infrastructure requirements for 6G will also need to improve significantly to meet these capabilities. As networks evolve, they will incorporate AI-driven optimization into their core, ensuring efficient resource allocation and dynamic network management. The integration of AI into 6G will create a self-learning, self-optimizing network that can adapt to MUs' needs in real-time. Among AI technologies, Deep Reinforcement Learning (DRL) offers some significant advantages in wireless communications. More specifically, DRL algorithms are effective in sequential decision-making environments, making them ideal for dynamic and uncertain systems. Moreover, DRL agents interact with the environment and learn from their experiences, thus it can achieve optimal long-term rewards under complicated, ever-changing conditions.

However, DRL is not without its drawbacks. Since DRL applies Deep Neural Network (DNN) for approximating the optimal policy, it needs a long time to learn the dynamics and uncertainty of the environment, especially in environments with large state spaces. Therefore, DRL can hardly meet the very low-latency requirements in wireless communications in highly dynamic communication systems where the environment changes quickly. For example, base stations in future communication systems are required to make radio resource allocation decisions to serve MUs in its coverage area within one transmission time interval which can be less than 1 ms. Therefore, running DRL with a rapidly changing environment and responding to new conditions within a transmission time interval is very challenging and difficult to achieve an optimal solution. Fortunately, the recent breakthrough in quantum computing has shown positive signs and may provide a potential solution due to its ability to significantly increase the computational speed compared to classical computation systems. This can be done due to the properties of quantum computing, such as quantum superposition and quantum entanglement. Quantum superposition allows a quantum bit (qubit) to exist in multiple states simultaneously until it is measured or observed. While classical computation systems store data in binary bits, where each of which can hold a value of 0 or 1. However, in quantum computing, a qubit can be in a state of 0, 1, or a combination of both states at the same time. This allows quantum computers to process multiple states at once, resulting in exponentially faster computations than traditional computers. Entanglement is explained as a phenomenon in which two or more particles are tightly bound and influence each other instantaneously, regardless of the huge physical distance between them. In quantum computing, changing the state of one qubit can immediately change another entangled qubit's state, thus significantly speeding up computation. Based on these properties, quantum reinforcement learning (QRL) has been proposed recently and shows promise to speed up the training of DRL. In QRL, conventional deep neural networks (DNNs) are replaced by variational quantum circuits or parameterized quantum circuits to approximate the optimal policy. Furthermore, QRL leverages quantum superposition and quantum entanglement to explore multiple states simultaneously, allowing for more efficient discovery of optimal policies, especially in environments with high-dimensional state spaces.


Although 6G has not yet been formally defined, we believe that QRL will become one of the key players. QRL will provide excellent computational capabilities to overcome the limitations of DRL, and bring many advantages in quantum-inspired security, quantum routing, quantum-inspired air-ground-sea networks, quantum-inspired real-time optimization, etc. Recently, some studies have proposed quantum machine learning (QML) for 6G and beyond networks \cite{wangQuantum6G,ZamanQuatum6GURLLC}. However, none of them have provided a comprehensive overview of QRL for 6G and beyond networks. They only discussed the general Quantum Machine Learning for 6G cellular networks \cite{wangQuantum6G} or for 6G ultra-reliable and low-latency communication (URLLC) \cite{ZamanQuatum6GURLLC}. The novelty and contributions of our work are as follows:

\begin{itemize}
    \item To the best of our knowledge, this article is the first one that discusses the fundamentals, applications, and case studies of QRL for 6G. 
    \item This article presents specific QRL opportunities for 6G. These opportunities include quantum routing, quantum space-air-ground-sea networks, quantum-inspired metaverse, quantum-inspired integrated sensing and communucations, quantum-inspired security, quantum-inspired real-time digital twins, and quantum-inspired holographic communications. Such information inspires researchers who extend their research to QRL-enabled 6G communication systems.
    \item This article provides some simulation results about the QRL-based dynamic spectrum access case study, a hot issue for 6G with limited spectrum and the increasing demand for capacity and data rate of users.
    \item Finally, we present a list of crucial challenges and future directions for QRL in 6G, such as 6G security, space-air-ground-sea communications, routing and massive access, ultra-reliable and low-latency communications, and integrated emerging technologies.
\end{itemize}

The rest of our work is organized as follows. First, we provide an overview of DRL for 6G in section \ref{section:DRL6G}. Second, we discuss the motivation, overview and applications of QRL for 6G in sections \ref{sec:qrl_motivation_foundations}, \ref{section:QRL6G}, \ref{section:QRLapp} respectively. Third, we provide a case study about QRL-inspired dynamic spectrum access in section \ref{section:QRLspectrum}. Finally, section \ref{section:conclusion} concludes our work.

\section{Overview of DRL}
\label{section:DRL6G}

A typical DRL framework consists of several key components: one or more \textbf{agents} interact with the environment and learn to make decisions that maximize their objectives (e.g., a base station with the objective of maximizing its throughput); \textbf{state space} ($S$), representing the agent’s perception of the environment (e.g., network traffic condition and channel information); \textbf{action space} ($A$), consisting of decisions that can be selected by the agent after observing environment states (e.g., bandwidth allocation and data routing decisions); \textbf{reward function} ($R$), which provides feedback on the effectiveness of the agents’ actions, guides the learning process by reinforcing behaviors that optimize long-term objectives (e.g., minimizing latency and maximizing throughput); and \textbf{policy} ($\pi$) defines the agent’s decision-making strategy, determining which action to take in a given state. In DRL, policies are often approximated using DNNs, enabling agents to generalize across complex environments.

As shown in \cite{hoang2023DRL4comms}, DRL is a versatile tool widely applied across wireless communication layers, with various architectures tackling key challenges in the field. At the physical layer, DRL is extensively applied to complex problems, including beamforming, signal detection, channel estimation, and channel decoding. For instance, in beamforming problems, diverse DRL techniques are leveraged, such as deep deterministic policy gradient (DDPG), multi-agent DRL, and approaches integrating model-based optimization. These techniques demonstrate improved performance in terms of spectral efficiency, bit error rate, convergence rate, and system capacity across various scenarios, including multi-input single-output, cell-free, RIS-aided, and mmWave systems.

Moving up the stack, at the medium access control (MAC) layer, DRL is applied to address complex problems like resource management and optimization, channel access control, and the coexistence of heterogeneous MAC protocols. Key DRL techniques used include deep Q-learning (DQN), double DQN, DDPQ, and multi-agent DDPG. DRL solutions have demonstrated significant improvements such as higher throughput, goodput, and fairness, lower latency, enhanced spectral efficiency, and better spectrum utilization. The applying scenarios include 5G radio resource scheduling, random access in massive IoT, power and rate control, and spectrum sharing.

At the network layer, DRL effectively manages complex challenges, particularly in traffic routing, network slicing, and network intrusion detection. DRL enables learning agents to optimize traffic flows, orchestrate virtualized resources, or identify security threats. For this class of problems, earlier mentioned techniques like DDPG, DQN and its variants are most broadly leveraged. Also, other promising techniques are also introduced, such as asynchronous advantage actor-critic, federated DRL, graph attention networks, and adversarial RL. DRL applications have shown benefits such as reduced delay, enhanced network utility, improved scalability, faster convergence, higher resource utilization, and superior performance against dynamic conditions, and often higher detection accuracy and robustness in intrusion detection.

Finally, at the application and service layer, DRL optimizes tasks like content caching, computation offloading, and data analytics to improve user experience and system efficiency under dynamic demands and resource constraints. DRL agents (e.g., DQN, DDPG) enable smart decisions on caching, offloading, and scheduling, often enhanced by techniques like prioritized replay, double DQN, and dueling networks. Benefits include higher cache hit rates, lower energy use, reduced delay, faster convergence, and improved QoS. Applications span IoT, virtual reality networks, mobile edge computing, fog computing, cloud databases, and distributed systems.

Although DRL has received significant attention in the literature and offers promising solutions across various areas of wireless communication, it still faces several challenges in future communication systems.

\textit{Challenge 1 - The complexity of network environments:} Modern communications networks involve a massive number of interconnected devices and fluctuating traffic patterns, which create a highly complex and ever-changing environment. Research studies on DRL techniques for handling wireless communications and networking problems mostly test their systems in very idealized and state-limited environments. More broadly, the dynamics of communications systems usually introduce significant complexity and uncertainty, making them difficult to be modeled or predicted. This complexity, as a result, presents a major challenge to the direct application and generalization of DRL-based solutions developed under simplified assumptions.

\textit{Challenge 2 - Long learning time:} DRL must overcome the challenge of system delays to make timely and effective decisions. In highly dynamic environments, the state may change by the time a decision is made based on observed information, resulting in outdated or misleading feedback that hampers the learning process. As a result, collecting reliable training data becomes particularly challenging, which in turn prolongs the overall DRL training process. Additionally, it is common in networking and communication scenarios where multiple agents operate in the same environment (e.g., spectrum control at base stations). In such scenarios, a DRL agent’s actions can alter the environment as perceived by other agents. This issue, known as nonstationarity \cite{hoang2023DRL4comms}, hinders policy convergence, making stable learning difficult.

\textit{Challenge 3 - The complexity of the DNN architecture:} To date, DNNs are often regarded as black boxes. Their outputs are highly dependent on the setting of learning parameters, such as the number of layers, activation functions, and weights of the nodes, rather than on transparent mathematical proof. Therefore, the implementation of DRL requires engineers with rich experience in both deep learning and wireless networks. Also, as mentioned earlier, DNNs used for decision-making in communication and networking problems require vast amounts of training data. This training process is highly resource-intensive, often exceeding the limited energy and processing capabilities of wireless devices.

\section{Motivations and Quantum Computing Fundamentals for QRL}
\label{sec:qrl_motivation_foundations}

Facing the aforementioned challenges of classical DRL in complex, large-scale environments like future 6G networks, researchers are exploring Quantum Reinforcement Learning (QRL) as a potentially powerful alternative. QRL aims to harness unique phenomena from quantum mechanics to enhance learning capabilities \cite{nielsen2010quantum, meyer2022survey}. This section outlines the motivation for exploring QRL and provides an intuitive introduction to the core quantum concepts involved.

\subsection{Why Consider Quantum Reinforcement Learning?}
\label{subsec:qrl_motivation_detailed}

The potential advantages of QRL stem directly from the counter-intuitive principles of quantum mechanics, which offer fundamentally different ways to represent and process information compared to classical computers :

\begin{itemize}
    \item \textbf{Exponential Representational Capacity:} Classical bits store either 0 or 1. Quantum bits, or qubits, can exist in superposition, representing both 0 and 1 simultaneously. This means $n$ qubits can represent $2^n$ classical states concurrently within a complex vector space called the Hilbert space \cite{nielsen2010quantum}. Imagine a classical library where each book must be read one by one. A quantum library of the same physical size could potentially hold an exponentially larger number of stories simultaneously accessible through superposition. This vast state space could allow QRL agents to potentially model highly complex environments or policies more compactly than DRL agents using DNNs \cite{meyer2022survey}.

    \item \textbf{Quantum Parallelism for Exploration:} Superposition allows quantum systems to perform computations on all components of their superimposed state simultaneously. In the context of RL, this characteristic, termed quantum parallelism, could theoretically enable an agent to evaluate or explore multiple state-action trajectories in parallel. For example, to explore a maze, a classical agent typically evaluates paths sequentially. In contrast, a quantum agent, by virtue of superposition, could effectively process information pertaining to numerous paths concurrently, potentially accelerating the discovery of optimal strategies \cite{nielsen2010quantum}.

    \item \textbf{Entanglement for Capturing Complex Correlations:} Entanglement creates a profound connection between two or more qubits, linking their properties instantaneously regardless of distance \cite{nielsen2010quantum}. A measurement performed on one qubit of an entangled pair yields information that is instantaneously correlated with the state of the other qubit(s). Such correlations are stronger than any achievable in classical systems. Within QRL, entanglement offers a powerful mechanism for efficiently modeling intricate, potentially non-local dependencies among diverse state variables or environmental features, dependencies that might otherwise necessitate highly complex or deep neural network architectures in classical DRL \cite{soohyun2024quantumautonomous}.

\end{itemize}
The confluence of these quantum characteristics, which include expansive representational capabilities, inherent parallelism, the capacity to model intricate correlations via entanglement, and the potential for computational effects through interference, motivates the investigation of QRL. It is hypothesized that these features could translate into tangible improvements in learning performance, such as accelerated policy convergence, enhanced sample efficiency leading to reduced data requirements, or the discovery of superior solutions within complex problem domains when compared to classical DRL paradigms \cite{meyer2022survey}. Nonetheless, the practical materialization of such quantum-derived advantages is contingent upon continued advancements in quantum hardware and the co-development of robust QRL algorithms.

\subsection{Intuitive Introduction to Quantum Computing Concepts}
\label{subsec:qc_basics_intuitive}

In the following, we provide an intuitive introduction to important concepts of quantum computing relevant to QRL, including qubits, quantum gates, entanglement, and measurement.

\begin{itemize}
    \item \textbf{Qubits - Beyond 0 and 1:}
        A classical bit is like a light switch: either OFF (0) or ON (1). A qubit is more like a dimmer switch or, better yet, a spinning coin. While it's spinning, it's neither heads nor tails but a combination of both possibilities. Mathematically, we label the definite states corresponding to 0 and 1 as $\state{0}$ and $\state{1}$ (read "ket 0" and "ket 1" - the $\ket{\cdot}$ notation is just a standard quantum label for a state). The qubit's state while "spinning" is a superposition, written as $\alpha\state{0} + \beta\state{1}$. Here, $\alpha$ and $\beta$ are complex numbers (probability amplitudes) that tell us the probability of getting 0 or 1 if we stop the coin (measure it). The key rule is $|\alpha|^2 + |\beta|^2 = 1$, meaning the probabilities must sum to 100\% \cite{nielsen2010quantum}. This superposition ability is what allows $n$ qubits to represent $2^n$ states simultaneously.

    \item \textbf{Quantum Gates (Manipulating Qubits):}
        Similar to how classical computers utilize logic gates (e.g., NOT, AND) to process bits, quantum computers employ quantum gates to manipulate the states of qubits \cite{nielsen2010quantum}. These are precisely controlled physical operations, such as the application of laser or microwave pulses, which coherently evolve the qubit's state. Such operations effectively alter the probability amplitudes, $\alpha$ and $\beta$, of the qubit's superposition.
        \begin{itemize}
            \item \textit{Single-Qubit Gates (Rotations):} These gates act on one qubit at a time. Prominent examples include the rotation gates $R_x(\theta)$, $R_y(\theta)$, and $R_z(\theta)$; these gates rotate the qubit's state vector about the x, y, or z-axis of the Bloch sphere (a geometrical representation of a single qubit's state space), respectively, by a specified angle $\theta$. In the context of QRL, parameterized quantum gates are frequently utilized, where the angle $\theta$ serves as a tunable variable. These angles are optimized during the learning process, functioning analogously to the adjustable weights in a classical neural network \cite{wang2025reinforcementlearningquantumcircuit}.
            \item \textit{Multi-Qubit Gates (Interactions):} These gates act on two or more qubits, making them interact. The most famous is the CNOT (Controlled-NOT) gate. It works on two qubits, a ``control'' and a ``target''. If the control qubit is in the $\state{1}$ state, the CNOT gate flips the target qubit (like a classical NOT gate: $\state{0} \to \state{1}$, $\state{1} \to \state{0}$). If the control qubit is $\state{0}$, it does nothing to the target. This conditional logic is crucial for building complex quantum algorithms and is the primary way to create entanglement between qubits \cite{nielsen2010quantum}.
        \end{itemize}

    \item \textbf{Entanglement:}
        Consider two qubits prepared in a specific correlated state, for instance, through the application of a CNOT gate. Entanglement is a uniquely quantum mechanical phenomenon where these qubits become linked in such a way that their individual states can no longer be described independently of each other; they share a single, joint quantum state \cite{nielsen2010quantum}. Consequently, upon measuring one qubit of an entangled pair and obtaining a specific outcome (e.g., $\state{0}$), the state of the other qubit is instantaneously determined (e.g., to be $\state{1}$ in a common entangled pair), irrespective of the physical distance separating them. This non-local correlation, famously characterized by Einstein as ``spooky action at a distance," enables entangled qubits to embody correlations far stronger and more complex than any achievable with classical bits. In QRL, this property offers a powerful resource for modeling intricate relationships between different components of an environment's state or an agent's actions \cite{soohyun2024quantumautonomous}.

    \item \textbf{Measurement:}
        While the superposition state of a qubit is essential for quantum computation, extracting a definitive classical result requires a measurement operation \cite{nielsen2010quantum}. The act of measurement projects the qubit from its superposition into a specific classical state, either $\state{0}$ or $\state{1}$. This outcome is fundamentally probabilistic, with the likelihood of obtaining $\state{0}$ given by $|\alpha|^2$ and $\state{1}$ by $|\beta|^2$. The inherent randomness associated with quantum measurement distinguishes it from deterministic classical operations and can be strategically exploited for exploration mechanisms in RL \cite{meyer2022survey}. Furthermore, it is possible to measure specific physical properties, termed observables (e.g., the Pauli Z operator). Repeated measurements of an observable allow for the estimation of its expectation value (i.e., the average outcome). Such expectation values provide continuous outputs that are valuable in QRL for tasks like estimating Q-values or defining continuous control parameters \cite{wang2025reinforcementlearningquantumcircuit}.
\end{itemize}

\section{Overview of Quantum Reinforcement Learning}
\label{section:QRL6G}

\begin{figure}[!]
	\centering
	\includegraphics[scale=0.9]{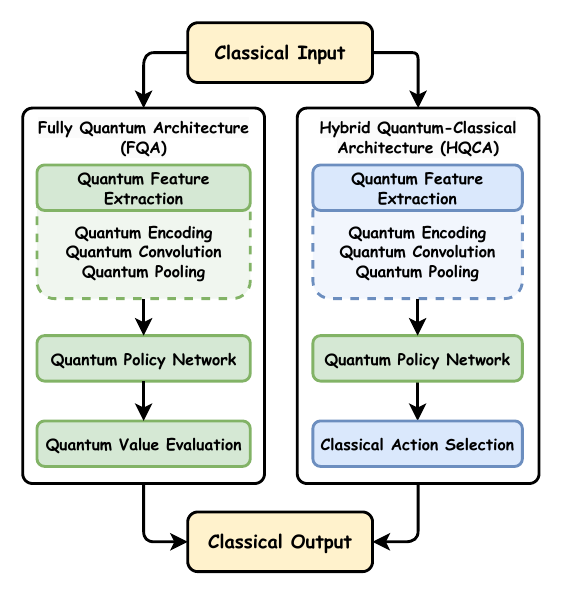}
	\caption{FQA and HQCA.}
	\label{fig:fqa&hqca}
\end{figure}

Quantum Reinforcement Learning (QRL) integrates the quantum computing principles previously outlined into the reinforcement learning framework, with the objective of enhancing operational performance, especially within the complex scenarios anticipated in 6G networks \cite{van2024dynamic, meyer2022survey}. Traditional Deep Reinforcement Learning (DRL) often encounters significant challenges related to scalability and convergence when applied to the demanding environments of 6G systems \cite{hoang2023DRL4comms}. In contrast, QRL aims to leverage quantum resources, such as superposition and entanglement, to potentially accelerate the learning process and manage vast state-action spaces with greater efficiency \cite{ van2024dynamic, nielsen2010quantum}. A pivotal aspect of many contemporary QRL strategies involves the replacement or augmentation of classical Deep Neural Network (DNN) components with Variational Quantum Circuits (VQCs) \cite{wang2025reinforcementlearningquantumcircuit}. The extent of this quantum integration fundamentally defines two primary architectural paradigms, which are illustrated in Fig.~\ref{fig:fqa&hqca} and draw inspiration from frameworks such as that introduced by Zaman et al. \cite{zaman2024comparativeanalysishybridquantumclassical}: Fully Quantum Architecture (FQA) and Hybrid Quantum-Classical Architecture (HQCA).

In FQA, quantum gates and measurements are employed to construct the entire RL framework, from feature extraction to decision-making. This approach leverages quantum parallelism and interference to explore state spaces efficiently, potentially leading to faster convergence than classical DRL. However, FQA’s feasibility hinges on the availability of large-scale quantum hardware and robust error correction techniques, both of which remain under development. Moreover, FQA is generally more suited to smaller-scale problems due to limited qubit availability, as the computational resources required to simulate large quantum circuits grow exponentially with the number of qubits.

Conversely, HQCA incorporates quantum circuits at specific stages of the RL pipeline, such as feature extraction or policy evaluation, while relying on classical neural networks for other tasks. This hybrid approach reduces the demand on quantum resources and mitigates noise issues inherent in current noisy intermediate-scale quantum (NISQ) devices. Although HQCA may not fully realize the theoretical maximum ``quantum speedup", it is more practical for today’s NISQ hardware, making it a feasible option for realworld applications. By exploiting quantum effects like superposition and entanglement, HQCA can provide more efficient feature representations, improving model performance without overburdening quantum resources. Classical systems, meanwhile, handle tasks where quantum advantages are less pronounced, ensuring scalability and practicality for complex problems \cite{zaman2024comparativeanalysishybridquantumclassical}.

Compared to traditional DRL, QRL presents several potential advantages. Quantum superposition and quantum entanglement enable a more efficient, parallel exploration of state-action spaces, which could accelerate convergence. Additionally, quantum states can represent information more compactly, reducing the number of parameters and computational requirements. Nevertheless, several technical challenges still exist. For FQA, limitations in quantum hardware availability and the exponential computational cost of simulating large-scale quantum circuits restrict its applicability. For HQCA, while it lowers hardware demands, it may not fully harness the speedup or expressive power of a fully quantum approach. The limited qubit count and noise in current quantum devices further complicate the implementation of large-scale, high-accuracy quantum circuits.

\begin{figure}[t!]
	\centering
	\includegraphics[scale=0.33]{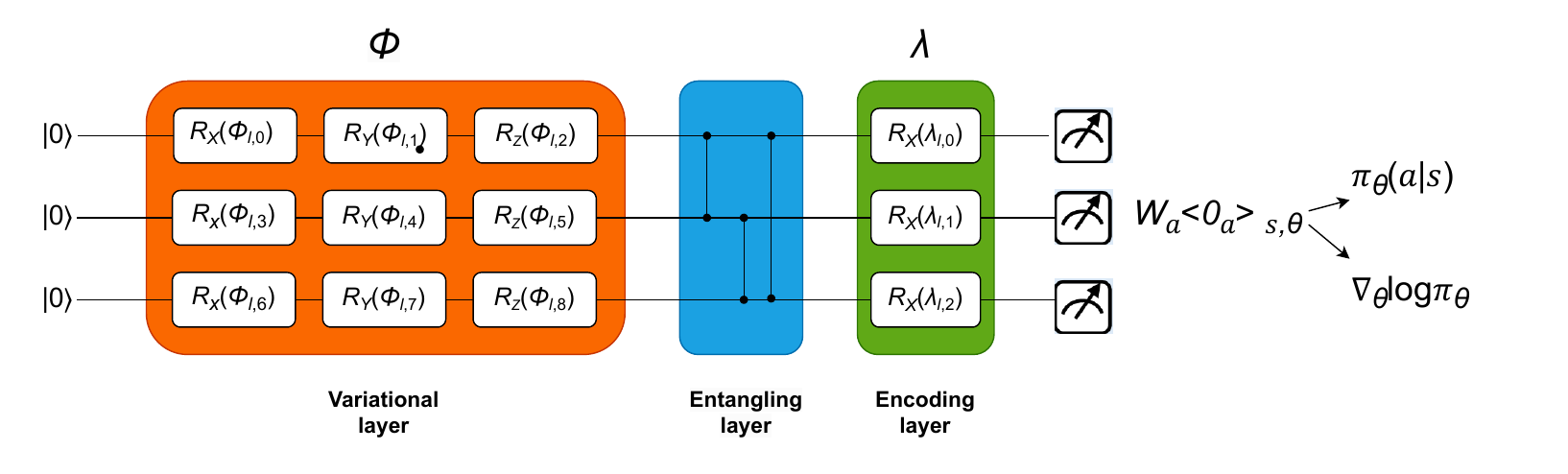}
	\caption{Architecture of a quantum layer.}
	\label{fig:quantum_layer}
\end{figure}

In the QRL workflow, qubits are typically initialized in the $\state{0}^{\otimes n}$ state, representing $n$ qubits each in the $\state{0}$ state. The quantum policy circuit, central to the agent's decision-making, is then constructed by applying a sequence of layered blocks. The structure of such a representative block, often visualized as in Fig.~\ref{fig:quantum_layer}, typically comprises three distinct stages applied in a specific order .
The first stage is a variational layer (commonly depicted in orange in diagrams). This layer applies parameterized single qubit rotation gates—such as $R_x(\phi_i)$, $R_y(\phi_j)$, and $R_z(\phi_k)$—to the individual qubits. The angles $\bm{\phi} = \{\phi_i, \phi_j, \phi_k, \dots\}$ within these gates constitute a set of tunable parameters. These parameters are optimized during the QRL training process, for instance, using gradient-based methods like the parameter shift rule. Following this, an entangling layer (blue) employs multi-qubit gates (such as CNOT or CZ) to generate nonlocal correlations (entanglement) between qubits. This is crucial for capturing complex dependencies within the environment's state space that purely local operations cannot. Finally, an encoding layer (green) re-introduces information about the current environment observation $s$. This is typically achieved through an angle-encoding scheme. In this scheme, features derived from the observation $s$ are used to determine the rotation angles, denoted by $\bm{\lambda} = \{\lambda_l, \lambda_m, \dots\}$, for another set of parameterized rotation gates (e.g., $R_x(\lambda_l(s))$, $R_y(\lambda_m(s))$). The parameters $\bm{\lambda}$ are thus functions of the input state $s$. This step effectively implements a data re-uploading mechanism within each block, allowing the circuit to process features of $s$ at multiple points in its depth. This entire block structure (Variational-Entangling-Encoding) can be repeated multiple times to increase the circuit's depth and representational capacity. 
By leveraging the high-dimensional Hilbert space and the interplay of parameterized rotations, entanglement, and repeated data injection, the circuit can learn a sophisticated and potentially more compact mapping $s \mapsto a$ compared to many classical deep networks. \cite{van2024dynamic}

Upon exiting the final layer, a measurement (also termed an observer) converts the quantum state into classical outputs for action selection. As highlighted in Fig.~\ref{fig:quantum_layer}, measuring in the computational basis yields bitstrings whose probabilities form the discrete policy $\pi(a \mid s)$. Alternatively, the agent may measure the expectation values of specific observables (e.g., $\langle Z\rangle$, or a linear combination of Pauli operators), providing continuous outputs useful for continuous control tasks or Q-value estimation. This innate measurement randomness offers built-in exploration, reducing the reliance on artificial noise or $\epsilon$-greedy heuristics. Meanwhile, the entanglement generated in earlier layers preserves global correlations across qubits, allowing the policy circuit to encode interactions that might demand numerous layers in a purely classical model. As the RL algorithm (e.g., policy gradient) incrementally updates the parameters $\phi$ and $\lambda$, the quantum circuit refines its decision strategy in tandem. Although current NISQ devices limit the circuit’s depth and qubit count, initial studies suggest that even small-scale quantum circuits, capitalizing on data re-uploading and entanglement, can rival or exceed classical baselines on certain benchmark tasks, highlighting the potential for improved sample efficiency and faster convergence in more advanced quantum hardware scenarios. \cite{van2024dynamic}. Besides this structure, several specific quantum circuit architectures relevant to QRL have been explored in the literature.

\begin{figure}[t!] 
	\centering
	\includegraphics[scale=0.6]{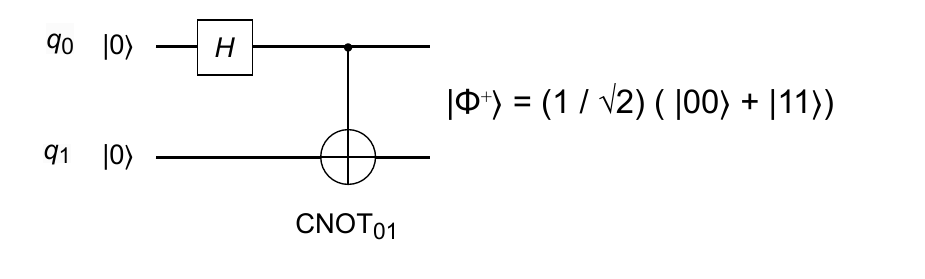} 
	\caption{A quantum circuit to generate the Bell state \(|\Phi^+\rangle\), illustrating elementary gate operations (inspired by matrix-based approaches 
\cite{wang2025reinforcementlearningquantumcircuit}).}
	\label{fig:quantum_layer_generateBell}
\end{figure}

In \cite{wang2025reinforcementlearningquantumcircuit}, Wang et al. propose a framework that uses a matrix-based approach to design quantum circuits via reinforcement learning (see Fig.~\ref{fig:quantum_layer_generateBell} for related concepts). This approach encodes the quantum state as a unitary matrix and treats elementary gates as matrix-based actions. It achieves high-fidelity synthesis of complex gates (e.g., iSWAP, CZ, Toffoli) and allows for transparent state evolution tracking via
\(
S' = A \cdot S.
\)
Techniques such as QR decomposition, givens rotations, and a reverse matrix method facilitate circuit construction, while discrete-action reinforcement learning algorithms (e.g., Q-learning, DQN) can effectively navigate the action space.
However, the approach scales poorly, as the state space dimension grows exponentially with the number of qubits, and each new target matrix requires separate training. It also exhibits moderate sampling efficiency and limited generalizability, especially in policy-based RL contexts. Consequently, while it is well-suited for precise gate design on small-scale simulators, it benefits from combining with methods such as Variational Circuits or Quantum Architecture Search for broader, more scalable applications..

\begin{figure}[t!]
	\centering
	\includegraphics[scale=0.75]{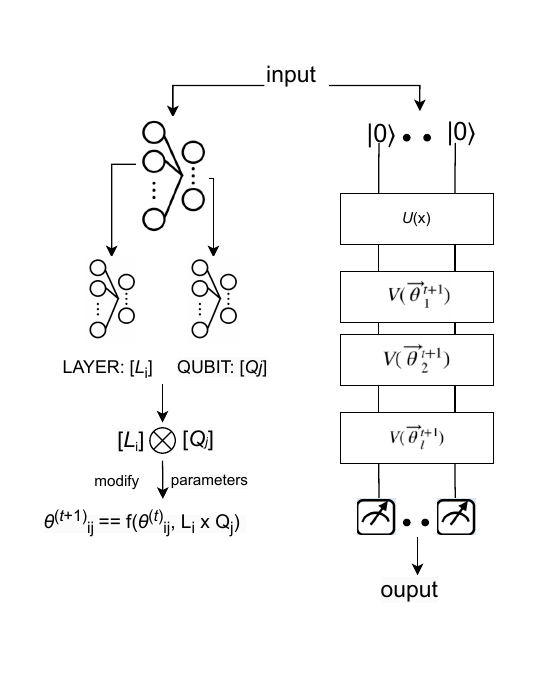}
	\caption{Quantum fast weight programmers.}
	\label{fig:quantumfwp_layer}
\end{figure}

As illustrated in Fig.~\ref{fig:quantumfwp_layer}, Chen et al. introduce the Quantum Fast–Weight Programmer (QFWP), a hybrid quantum-classical framework \cite{chen2024learningprogramvariationalquantum }. In this architecture,  a classical “slow programmer” (e.g., an LSTM or Transformer) generates incremental parameter updates for a “fast programmer,” realized as a Variational Quantum Circuit (VQC). By storing temporal information within the VQC parameters rather than a separate hidden state, QFWP avoids backpropagation-through-time in the quantum network and achieves faster convergence than quantum LSTM while also reducing trainable parameters by 70–90$\%$. It proves well-suited for near-term devices, i.e., Noisy Intermediate-Scale Quantum (NISQ), requiring only four to six qubits and quantum depths under ten layers. Despite these advantages, QFWP’s performance hinges on the classical model’s capacity; scalability degrades when qubit counts exceed ten, and concurrent learning rate tuning can lead to gradient vanishing or exploding. Additionally, its efficacy in multi-agent reinforcement learning or continuous action spaces remains largely unverified.

\section{Applications of QRL for 6G}
\label{section:QRLapp}
\begin{figure*}[t]
    \centering
    \includegraphics[scale=1.2]{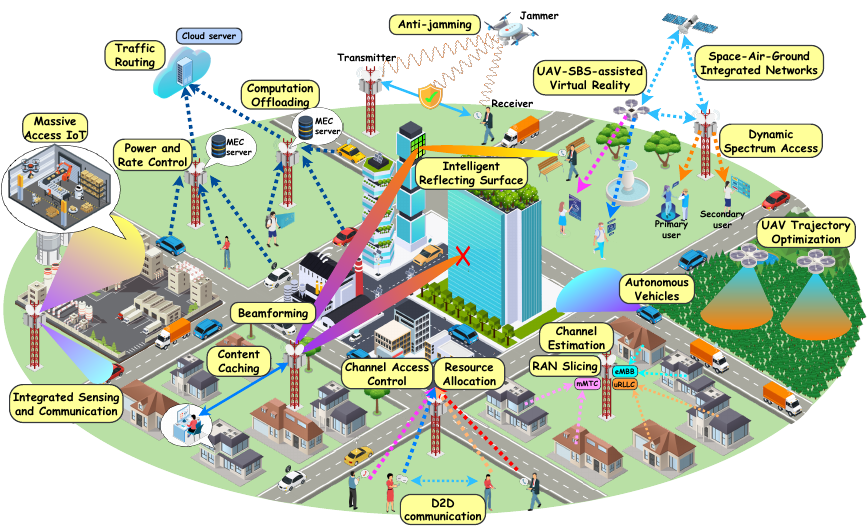}
    \caption{Applications of QRL in 6G.}
    \label{fig:DRL_applications}
\end{figure*}

As 6G evolve to meet demands for ultra-low latency, high data rates, and massive connectivity, key research areas have emerged, including dynamic resource allocation, UAV trajectory planning, multi-agent network control, and space-air-ground integrated networks (SAGIN). Addressing these challenges is vital for improving spectral efficiency, minimizing energy consumption, and ensuring reliable connectivity across diverse devices and services. Given QRL’s capability to handle large-scale optimization and decision-making tasks, it is well-positioned to address these issues in various 6G applications, as illustrated in Fig.~\ref{fig:DRL_applications}. The following sections explore how QRL applies to each area, highlighting recent advancements and outlining open challenges for future research.

\subsection{Dynamic Resource Allocation} 
Dynamic resource allocation stands out as a key area where QRL demonstrates considerable promise.  Within this context, quantum computing can be applied to optimize a variety of factors, including bandwidth distribution, power control, and channel assignment strategies. In non-orthogonal multiple access systems, QRL techniques enable the continuous adaptation of user pairings and power levels, thereby boosting spectral efficiency while maintaining QoS in mobile edge computing (MEC)-based IoT networks. Furthermore, for channel selection, Nuuman et al. in~\cite{nuuman2015quantuminspired} employ a quantum-amplitude table to iteratively select and refine channel assignments based on feedback regarding transmission success.  This approach has been shown to improve system capacity by 9\% to 84\% in terms of user count while concurrently reducing average file delay by approximately 26\% and increasing throughput by up to 2.8\%.


\subsection{Trajectory Planning for UAVs} 
The effectiveness of QRL has also been demonstrated in trajectory planning for unmanned aerial vehicles (UAVs) involved in tasks such as data gathering, coverage extension, and surveillance. In the context of multi-UAV systems, Park et al. in~\cite{soohyun2024quantumautonomous} propose a quantum multi-agent reinforcement learning (QMARL) framework. This framework integrates a quantum actor-critic design with centralized policy entanglement. Parameterized variational circuits are utilized to reduce the dimensionality of the action spaces, thus lowering the total number of trainable parameters. By adopting distributed quantum gradient synchronization, this QMARL approach achieves superior collision avoidance and reduced decision latency in dense UAV swarms. It outperforms classical multi-agent reinforcement learning in both convergence rate and overall efficiency. 

\subsection{Spectrum Access Optimization} 

Another significant contribution of QRL in 6G lies in optimizing spectrum access across numerous devices. Through continuous learning from the surrounding environment, QRL agents can dynamically allocate and reuse spectrum, thereby minimizing interference while simultaneously maximizing data rates. Furthermore, QRL offers the potential to optimize the deployment of backscatter communication technologies, allowing devices to exploit reflected ambient RF signals to lower energy consumption. In \cite{van2024dynamic}, the authors utilize QRL to enable dynamic spectrum access for D2D users employing backscatter communication, demonstrating the feasibility of data transmission even within congested frequency bands.

\subsection{Task Offloading and Integrated Sensing}

Task offloading strategies also benefit significantly from QRL. By judiciously weighing factors such as CPU load, bandwidth availability, and latency constraints, QRL-enabled devices can make informed decisions about whether to process tasks locally or offload them to edge or cloud servers. This adaptive capability ensures efficient resource utilization and reduced processing delays. In \cite{paul2024doa}, Paul et al. apply QRL in military surveillance scenarios that combine reconfigurable intelligent surfaces (RIS), UAVs, and satellite links. Their approach aims to improve both sensing accuracy (e.g., direction of arrival estimation) and offloading latency while adhering to the stringent requirements of URLLC.

\subsection{Multi-Agent Network Control} 

Within multi-agent 6G, QRL can facilitate enhanced cooperation and coordination among a diverse collection of devices, enabling them to collaboratively manage resources, mitigate interference, and adapt to dynamic traffic demands. Chaudhary et al. \cite{chaudhary2024vrcs} propose a QRL-based solution designed to optimize mode selection and resource allocation within vehicle-road collaborative systems. Their solution incorporates simultaneously transmitting and reflecting reconfigurable intelligent surfaces (STAR-RIS) with the aim of enhancing capacity, QoS, and scalability. This work clearly demonstrates the significant contribution that QRL can make to addressing the evolving demands of connected transportation in 6G networks.

\subsection{Space-Air-Ground Integrated Networks} 

Finally, QRL is well-suited to address the inherent complexity of resource scheduling in SAGIN, which encompasses satellites, UAVs, and terrestrial base stations. By learning optimal data routing paths and managing constrained energy resources, QRL can maintain stringent QoS requirements. In \cite{kim2024sagin}, Kim et al. demonstrate that QMARL-based techniques can mitigate the curse of dimensionality often encountered in SAGIN resource management, leading to reported improvements in both energy efficiency and overall network performance. These advancements are paving the way for the large-scale deployment of SAGIN within 6G architectures.

In summary, QRL is at the forefront of emerging techniques for achieving holistic optimization of 6G networks. Through quantum-enhanced exploration, accelerated convergence, and superior handling of expansive state-action spaces, QRL-based methods hold the potential to significantly advance resource allocation, mobility management, and service provisioning across a diverse spectrum of 6G use cases.

\section{Case Study: Quantum RL-based Dynamic Spectrum Access}
\label{section:QRLspectrum}
In this section, we conduct a case study on dynamic spectrum access to illustrate the advantages of QRL compared to conventional DRL-based approaches in terms of both learning performance and running time. In particular, we consider a device-to-device (D2D)-enabled cellular network consisting of a base station and multiple UEs. In our considered network, UEs can leverage D2D communications to communicate with each other, which will be referred to as D2D nodes in this paper, without sending data through the base station. This will be a typical setting in 6G networks to efficiently utilize limited wireless spectrum for a massive number of wireless devices. Without the loss of generality, the spectrum resource is shared between cellular communications and D2D communications in a time-splitting manner~\cite{huang2022dynamic}. We then define $\alpha$ as the probability that a UE accesses the shared spectrum in each time slot. As discussed, D2D may access this shared spectrum at the same time slots as UEs. Unfortunately, D2D nodes may cause interference and disrupt the communications of UEs in such scenarios. As studied in~\cite{huang2022dynamic}, UEs that are near the base station may suffer less from D2D interference due to high received signal strength. Hence, we define $\beta$ as the probability that a UE is in the base station's protected area, i.e., the distance between the UE and the base station is short enough to be immune from D2D interference.

Given the above, it is clear that D2D nodes must find appropriate time slots for their transmissions to avoid collisions with other UEs. This is particularly challenging in 6G, where a massive number of wireless devices with diverse requirements and behaviors coexist. To overcome this problem, QRL is emerging as a promising solution that can achieve more robust performance than state-of-the-art approaches. For that, we first formulate the dynamic spectrum access problem of D2D nodes as an MDP problem as in the following.

\begin{itemize}
    \item \textit{State:} To efficiently capture the system's properties and UEs' behaviors, the system state space includes the action performed in the last time slot, the channel state in the last time slot, the location of the UE in the previous time slot, and the distance between the current D2D transmitter node to the base station.
    \item \textit{Action:} The action space consists of two actions, including (i) the D2D transmitter stays idle and (ii) the D2D transmitter accesses the shared spectrum and transmits information to the D2D receiver.
    \item \textit{Immediate Reward:} After performing an action, the agent will observe the communication rate between the D2D transmitter and the D2D receiver, depending on their locations and the state of other UEs. The D2D transmission rate is calculated based on the Shannon capacity using the bilistic path-loss model.
\end{itemize}

\begin{figure}[t!]
	\centering
	\includegraphics[scale=0.45]{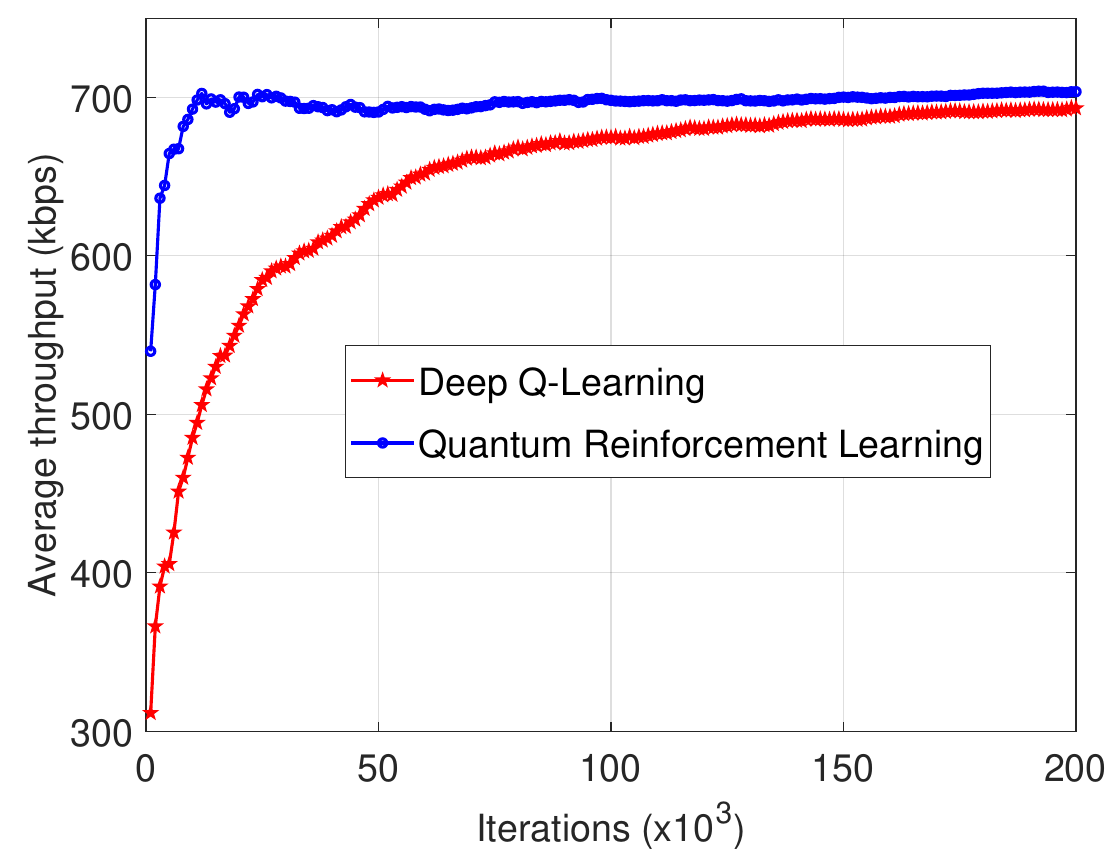}
	\caption{Convergence of quantum RL and DRL.}
	\label{fig:convergence}
\end{figure}
\begin{figure}[t!]
	\centering
	\includegraphics[scale=0.45]{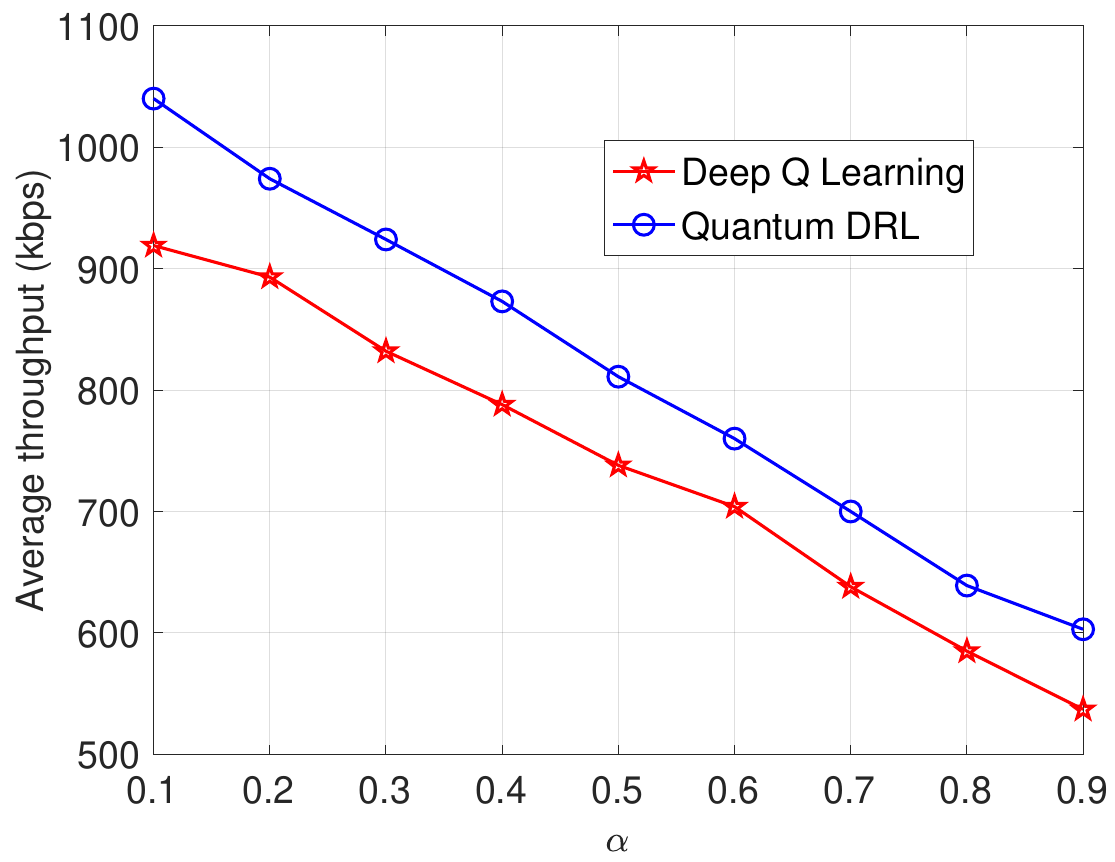}
	\caption{Average throughput vs. probability of UEs accessing the shared spectrum.}
	\label{fig:vary_p_access}
\end{figure}

Based on this Markov decision process formulation, we then implement the QRL algorithm and compare its performance with the conventional DRL algorithm with the following settings. Unless otherwise stated, the probability that a UE accesses the shared spectrum $\alpha$ is set at 0.7. The probability that a UE is in the protected area $\beta$ is set at 0.5. Without loss of generality, the distance between a D2D transmitter and the base station is randomly generated from 100 meters to 1,000 meters. The distance between a D2D transmitter and its receiver is also randomly generated from 20 meters to 100 meters. The transmit power of the base station and the D2D transmitter are set at 40 dBm and 23 dBm, respectively. The noise power is set at -114 dBm. The center frequency is set at 2 GHz, and the bandwidth is set at 20 MHz. For DRL, the DNN consists of one input layer, 2 hidden layers with a size of 64, and 1 output layer, making a total of 4610 training parameters. For the quantum circuit, we use 4 qubits to handle 4 features in our state space. In addition, 5 quantum layers, each consisting of one variational layer, one entangling layer, and one encoding layer, are designed to replace the conventional DNN. With this design, the quantum circuit has 94 trainable parameters, which is significantly smaller compared to that of the conventional DNN. 

In Fig.~\ref{fig:convergence}, we compare the convergence rate of QRL and conventional DRL algorithms. As can be observed, QRL can converge to the optimal policy much faster than DRL. Specifically, QRL can achieve an average throughput of 700 kbps after around 10,000 training iterations, while DRL requires more than 200,000 training iterations to obtain this throughput level. This demonstrates the effectiveness of using quantum circuits for decision-making problems, thanks to the quantum superposition principle. Additionally, in Fig.~\ref{fig:vary_p_access}, we vary the probability that UEs access the shared spectrum and evaluate the performance of QRL under different scenarios. Clearly, QRL performs better than DRL in all cases, confirming its great potential in 6G networks.

\section{Future Directions}
\label{section:futuredirection}

\subsection{6G Security}
The rapid development of quantum computing brings with it many technological breakthroughs but also some significant concerns. For example, quantum computers, with their powerful computing power, may make it easier for hackers to attack today's telecommunication networks. There has been a lot of interest from vendors and operators in quantum technologies for 6G security, e.g., IBM, AT$\&$T, Microsoft, Vodafone, Toshiba, and NTT. They are focusing on developing new technologies, such as Post-Quantum Cryptography, to counter hackers using quantum computers to break cryptography in current telecommunication systems. In addition, the investment in Quantum Key Distribution provides an additional layer of protection due to its ability to exploit the laws of quantum mechanics to create unbreakable encryption keys.

Besides Post-Quantum Cryptography and Quantum Key Distribution, DRL has demonstrated its effectiveness in combating cyberattacks. More specifically, DRL is capable of handling highly complicated, dynamic, and high-dimensional cyber-defense problems. For example, RL agents (e.g., MUs or base stations) choose different policies to improve the security of the wireless system. The learning agent will explore different rewards for each policy, thereby learning and finding the best policy under different environmental conditions. However, DRL also has the disadvantage of taking time to find a good policy for a fixed environment. Therefore, when the environment changes rapidly and has large state spaces, DRL will not be able to meet the requirements of applications that require very low latency, especially in 6G security. Fortunately, QRL will significantly improve the computation speed and help the algorithm run significantly faster than DRL and promises the potential to solve the existing disadvantages of DRL and dramatically improve 6G security.

\subsection{Space-Air-Ground-Sea Communications}
To ensure global connectivity, especially in areas where ground-based stations are not yet available, such as deserts, islands, forests, and offshore, Space-Air-Ground-Sea networks (SAGSNs) are considered a potential solution to compensate for the coverage and connectivity of terrestrial networks. SAGSNs not only include the distinct characteristics of each network (i.e., space, air, land, and sea) but also have the advantages of very wide coverage, heterogeneity, and guaranteed connectivity. Each individual network uses different access techniques and protocols. Therefore, choosing which technique or protocol to use in each network to ensure seamless communication brings new challenges. For example, ensuring continuous connectivity between Very Low Earth Orbit (VLEO) Satellite Systems and Maritime Systems while both systems move in separate orbits, especially as VLEO satellites move very fast, poses difficulties. Consequently, managing handovers between VLEO satellites or between VLEO and LEO/MEO/GEO satellites and applying beam hopping to meet throughput requirements for Maritime Systems while ensuring energy savings for satellites to increase their lifespan becomes a complex problem. Moreover, underwater communications often use acoustic communications, while satellite, ground, and air networks can use RF or optical communications. Therefore, ensuring continuous connectivity between asynchronous systems poses new challenges. Additionally, issues such as cross-layer design or routing and load-balancing in large-scale systems also present new challenges. As we know, DRL has been widely applied in space, air, terrestrial, and marine communications and responds to dynamic environments. However, with SAGSNs, DRL has to deal with large-scale networks with limited time constraints. Therefore, QRL significantly improves the computational capability of DRL, which will certainly bring new promises to SAGSNs. 

\subsection{Routing and Massive Access 6G}
Routing has always been a fundamental issue in communications. With billions of MUs, the need for routing connectivity to provide stable and continuous transmission becomes even more urgent in 6G. Especially in SAGSNs, the routing issue is to provide connectivity to isolated areas, such as ships using satellites or UAV communications. With a large number of satellites, the handover and routing between satellites to ground users to ensure continuity and throughput becomes very important. Especially when LEO/VLEO satellites have a short connection time to MUs, e.g., around ten minutes or less, and high speed of movement, continuous handover between satellites at different altitudes or many satellites cooperating to transmit data to MUs without interruption becomes crucial. The second problem is the trajectory design for UAV communications. With the high moving speed, the channel model between UAVs and MUs is constantly changing, not to mention obstacles and security problems. The design of UAVs' flight paths to ensure both MUs' data requirements as well as maintain the connection of UAVs to ground base stations and avoid flying into restricted areas becomes a challenge in finding the optimal flight path. Due to the countless flight points in the coverage area of UAVs, the UAV trajectory design problem is a non-convex, high-complexity problem. Third is routing in future radio access networks (RAN), e.g., Open RAN (O-RAN). O-RAN disaggregates a RAN block in the current system into different modules that can be developed and operated independently, for example, O-RAN Radio Unit (O-RU), O-RAN Distributed Unit (O-DU), O-RAN Central Unit User Plane (O-CU-UP), and O-RAN Central Unit Control Plane (O-CU-CP). Therefore, routing from MUs $\rightarrow$ O-RU $\rightarrow$ O-DU $\rightarrow$ O-CU-UP $\rightarrow$ Core networks with different capacity between each link, as well as dynamic traffic requirements and mobility of MUs, becomes a highly challenging problem. The fourth issue is ensuring connectivity between IoT devices in massive IoT networks. Billions of IoT devices are distributed in various services and applications such as autonomous vehicle communications, factory automation, traffic management, smart homes, agriculture, etc., with different requirements on latency and data volume. Therefore, designing routing strategies to ensure reliable wireless connectivity in massive IoT networks for data collection, especially in disaster and emergency cases, becomes crucial. These mentioned issues can easily become overwhelming with current computing systems. QRL is inspired by quantum computing with superior computing capabilities exponentially faster than current computing systems and the ability to find optimal routing policies will certainly bring superior results compared to conventional approaches.

\subsection{Ultra-Reliable and Low-Latency Communications}

Many applications, such as virtual reality, augmented reality, tactile communications, autonomous vehicles, live video streaming, and real-time optimization, have created huge challenges for URLLC in 6G. Especially when 6G promises to bring transmission rates up to 1 Terabit/second and latency can reach 0.1 ms, applying Deep Learning techniques such as DRL to O-RAN network slicing and resource allocation promises to bring many benefits. However, Deep Learning is challenged to meet URLLC requirements in highly dynamic environments. For example, its learning speed is reduced as the system dimensionality increases, thus potentially breaking the ultra-low latency requirement of 6G. Especially when the amount of data in 6G increases significantly with the majority being video traffic, it is very challenging for current methods to meet 6G URLLC requirements. By applying Quantum Machine Learning in general and Quantum Deep Learning in particular, replacing Variational Quantum Circuits or Parameterized Quantum Circuits for Deep Neural Networks (DNNs) to speed up achieving near-optimal policy, it has proven to be superior to classical DRL in solving complex optimization problems.

\subsection{Integrated Emerging Technologies}

In addition to the techniques discussed above, 6G will involve many other technologies such as intelligent reflective surfaces, integrated sensing and communications, holographic, tactile, and semantic communications, and digital twin networks. The integration of multiple emerging technologies aims to improve data rates, energy efficiency, spectrum efficiency, connection density, high mobility, and reliability. However, dealing with a large-scale network consisting of multiple emerging technologies with different optimizations also creates new challenges. Recently, hierarchical reinforcement learning has been applied to solve multiple optimization problems that run in different agents and are controlled by a master agent. This master agent decides which optimization problem to run at a particular time and in a particular area to optimize network performance under dynamically changing ambient conditions. As a result, optimization problems will become more complex and require more powerful computing resources to respond promptly. Quantum reinforcement learning (QRL) will become an effective tool to supplement hierarchical reinforcement learning. More specifically, we will have a QRL at each agent to handle separate optimizations and a QRL at the master agent to decide which global optimization will be used at a time to maximize network performance.

\section{Conclusion}
\label{section:conclusion}
In this article, we present a more comprehensive picture of QRL for 6G networks. More specifically, we first discuss the application challenges of DRL in 6G. Then, since these challenges cannot be solved in current communications systems due to their limitations, we discuss QRL and present current applications of QRL in 6G. Next, we discuss a use case of QRL in 6G, which is dynamic spectrum access; simulation results have demonstrated the superiority of QRL compared to DRL in terms of convergence rate and average throughput. Finally, we outline the potential directions of QRL for 6G, such as security, SAGSNs, routing and Massive Access 6G, URLLC, and integrated emerging technologies.

\bibliographystyle{IEEEtran}
\bibliography{IEEEfull}

\begin{thebibliography}{10}
\providecommand{\url}[1]{#1}
\csname url@samestyle\endcsname
\providecommand{\newblock}{\relax}
\providecommand{\bibinfo}[2]{#2}
\providecommand{\BIBentrySTDinterwordspacing}{\spaceskip=0pt\relax}
\providecommand{\BIBentryALTinterwordstretchfactor}{4}
\providecommand{\BIBentryALTinterwordspacing}{\spaceskip=\fontdimen2\font plus
\BIBentryALTinterwordstretchfactor\fontdimen3\font minus
  \fontdimen4\font\relax}
\providecommand{\BIBforeignlanguage}[2]{{%
\expandafter\ifx\csname l@#1\endcsname\relax
\typeout{** WARNING: IEEEtran.bst: No hyphenation pattern has been}%
\typeout{** loaded for the language `#1'. Using the pattern for}%
\typeout{** the default language instead.}%
\else
\language=\csname l@#1\endcsname
\fi
#2}}
\providecommand{\BIBdecl}{\relax}
\BIBdecl

\bibitem{wangQuantum6G}
C.~Wang and A.~Rahman, ``{Quantum-Enabled 6G Wireless Networks: Opportunities
  and Challenges},'' \emph{IEEE Wireless Communications}, vol.~29, no.~1, pp.
  58--69, 2022.

\bibitem{ZamanQuatum6GURLLC}
F.~Zaman, A.~Farooq, M.~A. Ullah, H.~Jung, H.~Shin, and M.~Z. Win, ``{Quantum
  Machine Intelligence for 6G URLLC},'' \emph{IEEE Wireless Communications},
  vol.~30, no.~2, pp. 22--30, 2023.

\bibitem{hoang2023DRL4comms}
D.~T. Hoang, N.~V. Huynh, D.~N. Nguyen, E.~Hossain, and D.~Niyato, \emph{Deep
  Reinforcement Learning and Its Applications}, 2023, pp. 1--23.

\bibitem{nielsen2010quantum}
M.~A. Nielsen and I.~L. Chuang, \emph{Quantum Computation and Quantum
  Information: 10th Anniversary Edition}.\hskip 1em plus 0.5em minus
  0.4em\relax Cambridge University Press, 2010.

\bibitem{meyer2022survey}
S.~Meyer, S.~Ahmed, M.~Sajid, I.~Ahmed, D.~Niyato, D.~I. Kim, and
  K.~Srinivasan, ``A survey on quantum reinforcement learning,'' \emph{arXiv
  preprint arXiv:2211.03423}, 2022.

\bibitem{soohyun2024quantumautonomous}
S.~Park, J.~P. Kim, C.~Park, S.~Jung, and J.~Kim, ``{Quantum Multi-Agent
  Reinforcement Learning for Autonomous Mobility Cooperation},'' \emph{IEEE
  Communications Magazine}, 2023.

\bibitem{wang2025reinforcementlearningquantumcircuit}
\BIBentryALTinterwordspacing
Z.~Wang, C.~Feng, C.~Poon, L.~Huang, X.~Zhao, Y.~Ma, T.~Fu, and X.-Y. Liu,
  ``{Reinforcement Learning for Quantum Circuit Design: Using Matrix
  Representations},'' 2025. [Online]. Available:
  \url{https://arxiv.org/abs/2501.16509}
\BIBentrySTDinterwordspacing

\bibitem{van2024dynamic}
N.~Van~Huynh, B.~Zhang, D.-H. Tran, D.~T. Hoang, D.~N. Nguyen, G.~Zheng,
  D.~Niyato, and Q.-V. Pham, ``{Dynamic Spectrum Access for Ambient Backscatter
  Communication-assisted D2D Systems with Quantum Reinforcement Learning},''
  \emph{arXiv preprint arXiv:2410.17971}, 2024.

\bibitem{zaman2024comparativeanalysishybridquantumclassical}
\BIBentryALTinterwordspacing
K.~Zaman, T.~Ahmed, M.~A. Hanif, A.~Marchisio, and M.~Shafique, ``A comparative
  analysis of hybrid-quantum classical neural networks,'' 2024. [Online].
  Available: \url{https://arxiv.org/abs/2402.10540}
\BIBentrySTDinterwordspacing

\bibitem{chen2024learningprogramvariationalquantum}
\BIBentryALTinterwordspacing
S.~Y.-C. Chen, ``{Learning to Program Variational Quantum Circuits with Fast
  Weights},'' 2024. [Online]. Available: \url{https://arxiv.org/abs/2402.17760}
\BIBentrySTDinterwordspacing

\bibitem{nuuman2015quantuminspired}
S.~Nuuman, D.~Grace, and T.~Clarke, ``{A Quantum Inspired Reinforcement
  Learning Technique for Beyond Next Generation Wireless Networks},'' in
  \emph{2015 IEEE Wireless Communications and Networking Conference Workshops
  (WCNCW)}.\hskip 1em plus 0.5em minus 0.4em\relax IEEE, 2015, pp. 271--275.

\bibitem{paul2024doa}
A.~Paul, K.~Singh, A.~Kaushik, C.-P. Li, O.~A. Dobre, M.~Di~Renzo, and T.~Q.
  Duong, ``{Quantum-Enhanced DRL Optimization for DOA Estimation and Task
  Offloading in ISAC Systems},'' \emph{IEEE Journal on Selected Areas in
  Communications}, 2024.

\bibitem{chaudhary2024vrcs}
S.~Chaudhary, I.~Budhiraja, R.~Chaudhary, N.~Kumar, D.~Garg, and A.~M.
  Almuhaideb, ``{Quantum Federated Reinforcement Learning Based Joint Mode
  Selection and Resource Allocation for STAR-RIS aided VRCS},'' \emph{IEEE
  Internet of Things Journal}, 2024.

\bibitem{kim2024sagin}
G.~S. Kim, Y.~Cho, J.~Chung, S.~Park, S.~Jung, Z.~Han, and J.~Kim, ``{Quantum
  Multi-Agent Reinforcement Learning for Cooperative Mobile Access in
  Space-Air-Ground Integrated Networks},'' \emph{arXiv preprint
  arXiv:2406.16994}, 2024.

\bibitem{huang2022dynamic}
J.~Huang, Y.~Yang, Z.~Gao, D.~He, and D.~W.~K. Ng, ``{Dynamic Spectrum Access
  for D2D-Enabled Internet of Things: A Deep Reinforcement Learning
  Approach},'' \emph{IEEE Internet of Things Journal}, vol.~9, no.~18, pp.
  17\,793--17\,807, 2022.

\end{thebibliography}

\begin{IEEEbiographynophoto}{Dinh-Hieu Tran} is currently a Research Associate at the University of Luxembourg. From 2022-2024, he was a Senior Research Specialist 5G+ at Nokia, France. He finished his M.Sc. (Hons.) at the Hongik University, Korea, in 2017, and his Ph.D. (Hons.) at the University of Luxembourg in 2022.
\end{IEEEbiographynophoto}

\begin{IEEEbiographynophoto}{Thai Duong Nguyen} is currently a Research Associate at the University of Engineering and Technology, Vietnam National University (VNU-UET). He finished his M.Sc at the VNU-UET in 2024. His research interests include AI $\&$ IoT, deep learning for wireless security, and 5G/6G. 
\end{IEEEbiographynophoto}

\begin{IEEEbiographynophoto}{Thanh-Dao Nguyen} is a Research Assistant of AI-CloudNetX laboratory at the University of Engineering and Technology, Vietnam National University (VNU-UET). His research interests include 5G and 6G communications, focusing on AI applications in wireless networks.
\end{IEEEbiographynophoto}

\begin{IEEEbiographynophoto}{Ngoc-Tan Nguyen} is currently a lecturer at the University of Engineering and Technology, Vietnam National University (VNU-UET). He received his Ph.D. degree at the University of Technology Sydney (UTS) in 2022. His research interests include Cloud/Edge computing, IoT, advanced wireless networks, 5G/6G, and AI/deep learning. 
\end{IEEEbiographynophoto}

\begin{IEEEbiographynophoto}{Van Nhan Vo} is currently a lecturer at Duy Tan University. He received his B.S., M.S., and Ph.D. degrees at the University of Da Nang and Duy Tan University, Vietnam, and Khon Kaen University, Thailand, in 2006, 2014, and 2019, respectively.
\end{IEEEbiographynophoto}

\begin{IEEEbiographynophoto}{Hung Tran} is the Head of the Data Technology and Communication Lab at the Faculty of Data Science and Artificial Intelligence, National Economics University, Vietnam. He earned his Ph.D. in 2013 from Blekinge Institute of Technology, Sweden. 
\end{IEEEbiographynophoto}

\begin{IEEEbiographynophoto}
{Mouhamad CHEHAITLY} received a Ph.D. degree (Hons.) at the Université de Lorraine, France, in 2017. From 2017 to 2020, he was an academic Teacher/Researcher at the Université de Grenoble Alpes and Université de Montpellier, with TIMA Laboratory, Grenoble, France. From 2020 to 2022, he conducted research at ISEN Engineering School, Brest, France. Since 2022, he has been a Research Associate at Luxembourg University with the SNT Laboratory.
\end{IEEEbiographynophoto}

\begin{IEEEbiographynophoto}
{Yan Kyaw Tun} received a Ph.D. degree in computer science and engineering from Kyung Hee University, Seoul, South Korea, in 2021. He is currently an Assistant Professor at the Department of Electronic Systems at Aalborg University, Denmark. 
\end{IEEEbiographynophoto}

\begin{IEEEbiographynophoto}
{Cedomir Stefanovic} received the Dipl.-Ing., Mr.-Ing., and Ph.D. degrees from the University of Novi Sad, Serbia. He is currently a Full Professor with the Department of Electronic Systems, Aalborg University, where he leads Edge Computing and Networking Research Group. 
\end{IEEEbiographynophoto}

\begin{IEEEbiographynophoto}
{Tu Dac Ho} is an Associate Professor in the Department of Information Security and Communication Technology, Norwegian University of Science and Technology (NTNU). 
\end{IEEEbiographynophoto}

\begin{IEEEbiographynophoto}
{Eva Lagunas} received PhD degree from the Polytechnic University of Catalonia (UPC), Barcelona, Spain, in 2014. In 2014, she joined the Interdisciplinary Centre for Security, Reliability and Trust (SnT), University of Luxembourg, where she currently holds a Research Scientist position. 
\end{IEEEbiographynophoto}

\begin{IEEEbiographynophoto}
{Symeon Chatzinotas} is currently Full Professor / Chief Scientist I and Head of the research group SIGCOM in the Interdisciplinary Centre for Security, Reliability and Trust, University of Luxembourg. In parallel, he is an Adjunct Professor at Norwegian University of Science and Technology, an Eminent Scholar of the Kyung Hee University, Korea, and a Collaborating Scholar of the Institute of Informatics \& Telecommunications, National Center for Scientific Research “Demokritos”. 
\end{IEEEbiographynophoto}

\begin{IEEEbiographynophoto}{Nguyen Van Huynh} is currently a Lecturer at the Department of Electrical Engineering and Electronics, University of Liverpool (UoL), United Kingdom. His research interests include cybersecurity, mobile computing, 5G/6G, IoT, and machine learning.
\end{IEEEbiographynophoto}

\vfill

\end{document}